# A Deep-Learning Approach for Operation of an Automated Realtime Flare Forecast

Yuko Hada-Muranushi,[1] Takayuki Muranushi,[2] Ayumi Asai,[1] Daisuke Okanohara,[3] Rudy Raymond,[3] Gentaro Watanabe,[3] Shigeru Nemoto,[4,5] and Kazunari Shibata[1]

**Abstract.** Automated forecasts serve important role in space weather science, by providing statistical insights to flare-trigger mechanisms, and by enabling tailor-made forecasts and high-frequency forecasts. We have been operating unmanned flare forecast service since August, 2015 that provides 24-hour-ahead forecast of solar flares, every 12 minutes. We report the method and prediction results of the system.

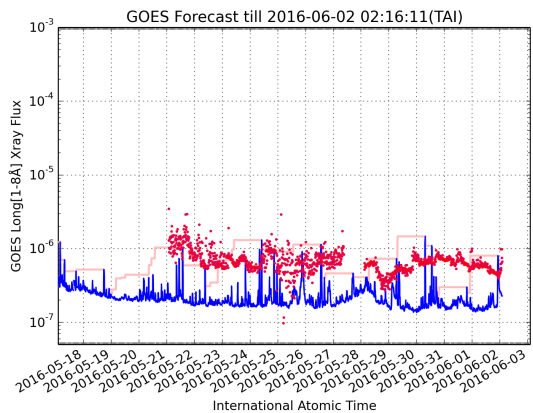

**Figure 1.** A screenshot from our forecast website. The blue curve is the observed Solar X-ray Flux (1-8 Å). The red dots are our forecast of the 24-hour future maxima of the Solar X-ray Flux. The pale red curve indicates the correct prediction, in retrospect. Our ideal goal is to have all the red dots on the pale red curve.

## 1. Introduction

Space weather impacts various aspects of human activities and thus space weather forecast is important to human society[*National Research Council*, 2008]. Among various space weather events, we focus on the forecast of the solar flares. This is because solar flares are the cause of the chain of space weather events that takes place in the interplanetary space and at Earth magnetosphere. Therefore, improvement of solar flare prediction improves both the accuracy and the lead time of the forecasts of the subsequent space weather events.

Machine learning is study of algorithms to construct models from known pairs of inputs and outputs, that can be used to predict outputs for unknown inputs. In space weather, there are decades of observational data of high-resolution images of the Sun and solar X-ray flux. Therefore, a lot of input-output data are available, making space weather a good application of machine learning.

Previously, various machine learning techniques have been applied to flare prediction algorithms: support vector machines (SVM) [*Li et al.*, 2007; *Bobra and Couvidat*, 2014; *Muranushi et al.*, 2015], ordinal logistic regression [*Song et al.*, 2009; *Yuan et al.*, 2010], neural networks [*Colak and Qahwaji*, 2009; *Yu et al.*, 2009; *Ahmed et al.*, 2013]. *Nishizuka et al.* [2016] compared the performance of SVM with those of the k-nearest neighbor (k-NN) and the extra random trees (ERT) algorithms.

Since 2010's, a new kind of machine learning technique called deep learning have been drawing attention by breaking the records in various machine learning competition. Deep learning is the study of how to train large-scale neural networks by applying stochastic gradient descent. Deep learning has two advantages for space weather applications.

One is that deep learning can accept much larger data as input features. Previous machine-learning algorithms such as SVM expected the size of input to be at most about $10^4$ real numbers, and each real numbers are meant to be *features*, representative parameters computed from raw data. The feature vectors have to be designed manually. In deep learning, raw data such as images or time series can be directly input to machine learning algorithms. Thus deep learning reduces the labor of designing feature vectors.

The other is that deep learning algorithms are *online learning* algorithms. Online learning algorithms are those that can incrementally learn as new data items become available. On the other hand, *batch learning* algorithms can learn only when all data are available. In real-time forecast applications, new data becomes continuously available and therefore online learning are favorable.

We have been operating automated, real-time flare forecast system using deep learning. The system is implemented using Chainer [*Tokui et al.*, 2015], a python-based programming framework for deep learning. Chainer adopts a "Define-by-Run" scheme, i.e. the network is defined on-the-fly via the actual forward computation. Users can specify flexible neural network archtectures using python codes.

The purposes of the automated forecast is twofold: One is to provide continuous, operational flare forecast. The other is to experimentally measure the performance of flare-forecasting methods, with the realtime forecast, avoiding any statistical flukes.

[1] Kwasan and Hida Observatories, Kyoto University, Yamashina-ku, Kyoto 607-8471, Japan
[2] RIKEN Advanced Institute for Computational Science, 7-1-26, Minatojima-minami-machi, Chuo-ku, Kobe, Hyogo, 650-0047, Japan
[3] Preferred Networks. Inc. Otemachi Bldg. 2F, 1-6-1, Otemachi, Chiyoda-ku, Tokyo, 100-0004, Japan
[4] Unit of Synergetic Studies for Space, Kyoto University, Kitashirakawa Oiwake-cho, Sakyo-ku, Kyoto 606-8502, Japan
[5] BroadBand Tower, Inc. 1-3-2 Uchisaiwai-cho, Chiyoda-ku, Tokyo 100-0011, Japan

The forecast is provided at http://www.spaceweather.kyoto/, a forecast and outreach website in Japanese. A more technical version of the forecast is provided at http://54.187.234.47/prediction-result.html, in English. An example of visualized prediction history is in Figure 1.

## 2. Design of The Realtime Forecast System

We use UFCORIN (Universal Forecast Constructor by Optimized Regression of Inputs). It can be used to predict general time series, based on the set of input time series. It is an open-source software published at https://github.com/nushio3/UFCORIN under MIT license.

We predict 24-hour future maximum of the GOES X-ray Flux (1-8 Å). Predicting $n$-hour future maximum of GOES X-ray light curve at time $t$ is equivalent to predicting the largest flare that will occur in the next $n$-hours.

Figure 2 illustrates our real-time forecast system. The system is executing the following operations continuously:

1. Obtain JSOC HMI image data, every 360 second.
2. Obtain GOES X-ray flux data, every 60 second.
3. Extract features from wavelet transformation using the method of [*Muranushi et al.*, 2015].
4. Store the feature in the database.
5. Repeatedly run the learning server.
6. Once in 720 second, execute the prediction server and generate the forecast.

We have used two major sources of observational data. One is the GOES Solar X-ray flux data [*Space Weather Prediction Center Website*, 2014] provided by National Oceanic and Atmospheric Administration (NOAA). The other is line-of-sight magnetic field image data from the Helioseismic and Magnetic Imager (HMI; *Scherrer et al.* [2012]) on board the Solar Dynamic Observatory (SDO; *Schou et al.* [2012]).

We use True Skill Statistics (TSS), as suggested by *Bloomfield et al.* [2012], to evaluate the predictions. TSS is calculated from the contingency table (c.f. Table 1), as follows:

$$TSS = \frac{TP}{TP+FN} - \frac{FP}{FP+TN}. \quad (1)$$

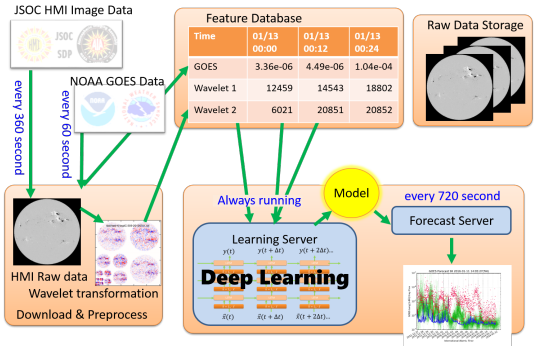

**Figure 2.** Overview of our realtime forecast website

**Table 1.** A contingency table, that counts true positive (TP), true negative (TN), false positive (FP), and false negative (FN) events.

| event | Observed | Not observed |
|---|---|---|
| Predicted "Yes" | $TP$ | $FP$ |
| Predicted "No" | $FN$ | $TN$ |

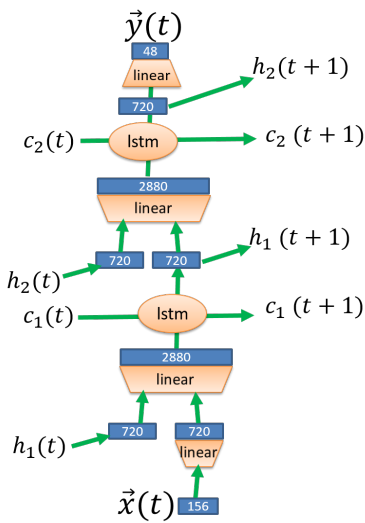

**Figure 3.** The neural network used for our forecast

## 3. The Forecast Neural Network

### 3.1. Overview of the forecast

The neural network we have designed takes 156 input parameters, and predicts 48 different output values, per 720-second cadence. The neural network have three hidden layers, and consists of several linear functions and two long-short term memory (LSTM) units.

### 3.2. The input

Our neural network takes 156 input data every 720 seconds. The 156-dimension vector of the input data at time $t$ is denoted by $\vec{x}(t)$. The ingredients of $\vec{x}(t)$ are as follows:

• 1 real number for the availability of the GOES X-ray flux data (1.0 when both long and short wavelength data are available, 0.0 otherwise.)

• 2 real numbers for the GOES X-ray flux data, one for the short $(0.5-4.0$ Å$)$ and one forthe long $(1.0-8.0$ Å$)$ wavelength. The maximum value of the corresponding 720-second span is used.

• 1 real number for the availability of the HMI line-of-sight data.

• 121 real numbers for the features constructed from the *standard*, 2d wavelet transformation of the HMI image using the Haar basis.

• 31 real numbers for the features constructed from the *non-standard*, 2d wavelet transformation of the HMI image using the Haar basis.

**Table 2.** Dimensions of the vectors, matrices and functions that appear in the prediction neural network.

| | |
|---:|:---|
| $\vec{x}$ | 156-dimensional vector |
| $L_0$ | $720 \times 156$ matrix |
| $\vec{h}_0, \vec{h}_1, \vec{h}_2$ | 720-dimensional vector |
| $L_{h1}, L_{x1}$ | $2880 \times 720$ matrix |
| $h_{1,in}, h_{2,in}$ | 2880-dimensional vector |
| lstm | a 2880-dimensional and a 720-dimensional vector to two 720-dimensional vectors |
| $\vec{c}_1, \vec{c}_2$ | 720-dimensional vector |
| $L_{h2}, L_{x2}$ | $2880 \times 720$ matrix |
| $L_3$ | $48 \times 720$ matrix |
| $\vec{y}$ | 48-dimensional vector |

The algorithm for computing wavelet features are the same as that in [*Muranushi et al.*, 2015].

### 3.3. The output

At every 720 seconds, $\vec{y}(t)$, the vector of 48 numbers that constitute our prediction are computed. The neural network is trained so that $\vec{y}(t)$ matches the future light curve $\vec{y}^{(\ell)}$ and the future maxima $\vec{y}^{(f)}$ in the limit of perfect prediction:

$$y_i(t) = y_i^{(\ell)}(t) \quad \text{for } 0 \leq i < 24, \quad (2)$$
$$y_i(t) = y_{i-24}^{(f)}(t) \quad \text{for } 24 \leq i < 48. \quad (3)$$

The target values $\vec{y}^{(\ell)}$ and $\vec{y}^{(f)}$ are defined as follows:

$$y_i^{(\ell)}(t) = \max_{t+iT < t' < t+(i+1)T} F_{\mathrm{X}}(t'), \quad (4)$$
$$y_i^{(f)}(t) = \max_{t < t' < t+(i+1)T} F_{\mathrm{X}}(t'). \quad (5)$$

for $0 \leq i < 24$. Here, $F_{\mathrm{X}}(t')$ is the GOES X-ray flux, and prediction cadence $T = 1$ hour.

The outputs $y_i^{(\ell)}(t)$ represents 1-hour maximum starting from 0 to 23 hours in the future. In other words, $y_i^{(\ell)}(t)$ constitutes GOES lightcurve forecast for the next 24 hours, with resolution of 1 hour.

The outputs $y_i^{(f)}(t)$ represents the forecast for the largest flare that will take place in the next $i$ hours in the future. The last element $y_{23}^{(\ell)}(t)$ is reported at the forecast website as the forecast for the next 24 hours.

### 3.4. The neural network

Our neural network computes $\vec{y}(t)$, the predictions at time $t$, from the observational data $\vec{x}(t)$, as follows:

$$\vec{h}_0 = L_0 \vec{x}(t) \quad (6)$$
$$\vec{h}_{1,\mathrm{in}} = L_{\mathrm{h}1} \vec{h}_1(t) + L_{\mathrm{x}1} \vec{h}_0 \quad (7)$$
$$(\vec{c}_1(t+1), \vec{h}_1(t+1)) = \mathrm{lstm}(c_1(t), \vec{h}_{1,\mathrm{in}}) \quad (8)$$
$$\vec{h}_{2,\mathrm{in}} = L_{\mathrm{h}2} \vec{h}_2(t) + L_{\mathrm{x}2} \vec{h}_1(t+1) \quad (9)$$
$$(\vec{c}_2(t+1), \vec{h}_2(t+1)) = \mathrm{lstm}(c_2(t), \vec{h}_{2,\mathrm{in}}) \quad (10)$$
$$\vec{y}(t) = L_3 \vec{h}_2(t+1) \quad (11)$$

The same computation is illustrated in Figure 3. The dimensions of the vectors and matrices in Equations (6-11) are summarized in Table 2.

In these equations, the symbols $L_0, L_{\mathrm{h}1}, \cdots$ represent linear functions from vector to vector, or matrix multiplications. The symbol lstm denotes the long-short term memory [*Hochreiter and Schmidhuber*, 1997] function, that takes a pair of $n$-dimensional vector and a $4n$-dimensional vector, and returns a pair of two $n$-dimensional vectors.

The action of the LSTM function, $(\vec{c}', \vec{h}') = \mathrm{lstm}(\vec{c}, \vec{h})$, is defined as follows:

$$c'_i = \varsigma(h_{4i}) h_{4i+3} + \varsigma(h_{4i+1}) c_i, \quad (12)$$
$$h'_i = \varsigma(h_{4i+2}) c'_i, \quad (13)$$

where $\varsigma$ is the sigmoid function:

$$\varsigma(x) = \frac{1}{1+\exp(-x)} \quad (14)$$

As is usual with deep learning, the free parameters of the prediction model are the elements of the matrices $L_0, L_{\mathrm{h}1}, \cdots$. These parameters are optimized in order to improve the output. We denote the set of parameter as $\vec{w}(t)$.

The model has 8,441,280 parameters in total. To optimize models with such a large number of degree of freedom while avoiding overfitting is the art of deep learning.

### 3.5. The learning method

In optimizing the parameters $\vec{w}(t)$ in the neural network, stochastic gradient descent methods are most commonly used. In the basic stochastic gradient descent method (hereafter SGD), $w_i(t+1)$, the value of the parameter after learning the $t$-th data is computed from $w_i(t)$ as follows:

$$w_i(t+1) = w_i(t) - \eta \frac{\partial L}{\partial w_i} \quad (15)$$

Here, $\eta$ is a small parameter (typically $\eta = 0.01$) called *learning rate*. Stochastic gradient descent is stochastic in sense that the outcome of algorithm depends on the order of the input data $x(t)$, that is drawn from model probability distribution. Once the data sequence $\{\vec{x}(t)\}$ is fixed, stochastic gradient descent algorithms are deterministic.

The convergence of the basic stochastic gradient descent algorithm is much worse if the function to optimize are noisy, or have a long valley. In order to avoid local noise and capture the global gradient in such a case, *momentum SGD* algorithm is used. The equations for momentum SGD are as follows:

$$v_i(t+1) = \alpha v_i(t) - \eta \frac{\partial L}{\partial w_i} \quad (16)$$
$$w_i(t+1) = w_i(t) + v_i(t+1) \quad (17)$$

Here, typically $\alpha = 0.9$ is used. The newly introduced variable $v_i(t)$ is the weighted sum of the gradient $\frac{\partial L}{\partial w_i}$ for the past several samples.

More sophisticated stochastic gradient descent algorithms are proposed, in aim to achieve better optimization, for example by dynamically changing the learning rate $\eta$. Examples of such algorithms are AdaGrad [*Duchi et al.*, 2011], RMSprop [*Tieleman and Hinton*, 2012], AdaDelta [*Zeiler*, 2012], and Adam [*Kingma and Ba*, 2014].

## 4. Simulated Forecast Experiments

### 4.1. Method

Before starting real-time forecast, we surveyed for best set of forecast parameters, or, the best prediction strategy.

**Table 3.** The list of surveyed parameters.

| backpropagation length | accel, 2, 32, 1024 |
|---|---|
| gradient factor | AB, flat, severe |
| optimizer | AdaDelta, AdaGrad, Adam, MomentumSGD, RMSprop, SGD |
| learning rate multiplier | 0.1, 0.3, 1, 3, 10 |

**Table 4.** The list of best-performing prediction strategies for each flare class, for the simulated forecast on the data of years 2011-2014.

| Flare class | X | ≥M | ≥C |
|---|---|---|---|
| Backprop length | 2 | 2 | 2 |
| grad factor | flat | severe | severe |
| optimizer | MomentumSGD | MomentumSGD | AdaDelta |
| multiplier | 10.0 | 3.0 | 10.0 |
| TSS | 0.736 | 0.671 | 0.635 |

The surveyed parameters and their range are as follows (c.f. Table 3):

- *Backpropagation length* is the number of steps when we perform backpropagation through time (BPTT). we have surveyed for 2, 32, 1024 steps, respectively, that corresponds to 24 minutes, 6.4 hours, and approximately 8.5 days, respectively. In addition, we have programmed the special backpropagation length parameter *accel*, where backpropagation length is initially set to 2, and is multiplied every 20 hours, until it reaches 1024.
- *Gradient factor* is the multiplier to the calculated gradient. We have compared three types of multiplier: AB, flat, and severe.
- *Optimizers* are stochastic gradient descent algorithms for training neural networks. We have compared the following six optimizers: AdaDelta [*Zeiler*, 2012], AdaGrad [*Duchi et al.*, 2011], Adam[*Kingma and Ba*, 2014], MomentumSGD, RMSprop[*Tieleman and Hinton*, 2012], and SGD.
- *Learning rate multiplier* This parameter controls the learning rate. Larger learning rate means the late-coming data have greater effect on the model.

The experiment was conducted on an 8-core machine using GNU Parallel [*Tange*, 2011].

### 4.2. Results

Table 4 lists the prediction strategies that achieved best TSS for X, $\geq$M, and $\geq$C class flares. Forced to choose one from the best three, we decided to focus on the prediction of the $\geq$M-class flares, since it is the class of flares that take place often enough and yet impose non-negligible effects on our space assets. Thus the following strategy is adopted for our forecast service, which presented the best TSS for $\geq$M class flare forecast.

- backpropagatin length:2
- gradient factor: severe
- optimizer: MomentumSGD
- learning rate multiplier: 3 ($\eta = 0.03$)

## 5. Operation of The Realtime Forecast

### 5.1. Method

The first prediction was made at 2015-08-13T 04:57 TAI. Since then, more than 25,000 predictions have been made. Figure 4 shows the result of the predictions made so far. We have experienced five measure incidents that caused the discontinuation of the prediction operation. The incidents with their reasons are listed in Table 7. As noted, we have learned from these incidents, have implemented countermeasures to prevent the similar kind of failure in the future.

### 5.2. Results

Table 5 shows the contingency table for our real-time forecast. No X-class flare have taken place in the operation period, so the contingency table for only $\geq$M-class and $\geq$C-class forecasts are included. TSS for $\geq$M and $\geq$C classes are 0.295 and 0.269, respectively.

**Table 5.** A contingency table of our real-time forecast for $\geq$M and $\geq$C class events since 2015-08-13T 04:57 TAI till 2016-06-02T 00:03 TAI.

| $\geq$M-class event | Observed | Not observed |
|---|---|---|
| Predicted | 1,178 | 2,555 |
| Not predicted | 1,873 | 19,264 |

| $\geq$C-class event | Observed | Not observed |
|---|---|---|
| Predicted | 10,332 | 4,516 |
| Not predicted | 3,968 | 6,054 |

## 6. Conclusions and discussions

UFCORIN have been participating CCMC Flare scoreboard [*Murray et al.*, 2015]. The scoreboard collects the automated flare forecasts from various institution around the world to foster international research community. The first submission by UFCORIN was made at 2016-05-30T 08:03 TAI.

In order to participate in the scoreboard, a forecaster is required to submit the probability for X, $\geq$M and $\geq$C class flares for the next 24 hours. Alternatively, the forecaster may choose to submit the probability for X, M and C class. Forecast for flare regions are optional. At the moment, we choose to submit only the full-disk forecast, for X, $\geq$M and $\geq$C class flares.

While the CCMC Flare scoreboard requires *probabilistic* forecast for the three *classes*, UFCORIN generates *deterministic* forecast for the Solar X-ray flux *values*. We must somehow convert the latter to the former in order to participate. Our current method is as follows.

First, we estimate $p(x)$, the probabilistic density function of the logarithm of the Solar X-ray flux maximum in the next 24 hours, by a normal distribution. We assume that the mean of the distribution $\mu = y_{47}$, the 24-hour future max forecast value. We estimate the standard deviation $\sigma$, by the maximum of $|y_{43} - y_{47}|, \cdots, |y_{46} - y_{47}|$, or, future max predictions whose span is no smaller than 20 hours.

$$p(x) = \mathcal{N}(x; \mu, \sigma), \quad (18)$$
$$\mu = \log y_{47}, \quad (19)$$
$$\sigma = \left\{ |\log y_i - \log y_{47}| \,\Big|\, i \in \{43, \cdots, 46\} \right\}. \quad (20)$$

Under this model, the cumulative density function $\mathrm{cdf}(x)$, or the probability of the 24-hour Solar X-ray flux maximum being greater than $x$, is as follows:

$$\mathrm{cdf}(x) = \int_x p(x')dx'$$
$$= \frac{1}{2}\left(1 - \mathrm{erf}(x; \mu, \sigma)\right). \quad (21)$$
$$\quad (22)$$

Using this $\mathrm{cdf}(x)$, the probabilistic forecast for X, $\geq$M, and $\geq$C class flares are calculated as follows:

$$P_{\mathrm{X}} = \mathrm{cdf}(\log 10^{-4}), \quad (23)$$
$$P_{\geq \mathrm{M}} = \mathrm{cdf}(\log 10^{-5}), \quad (24)$$
$$P_{\geq \mathrm{C}} = \mathrm{cdf}(\log 10^{-6}). \quad (25)$$
$$\quad (26)$$

We submit these values to CCMC flare scoreboard as UFCORIN's forecast.

## 7. Conclusions and discussions

TSS values of the real-time forecasts are much worse than those of the simulated forecast. This poses a serious issue on how to construct a good realtime forecast. Prediction strategy that performed well in the past does not necessarily shows the same performance in the realtime forecast.

**Table 6.** The list of Amazon Web Services (AWS) machines used for realtime forecast

| AWS | node type | CPU | Mem | Disk |
|---|---|---|---|---|
| RDS | db.m3.medium | 1 | 3.75 | 100 |
| EC2 | m3.large | 2 | 7.5 | 62 |
| EC2 | g2.2xlarge | 8 | 15 | 123 |

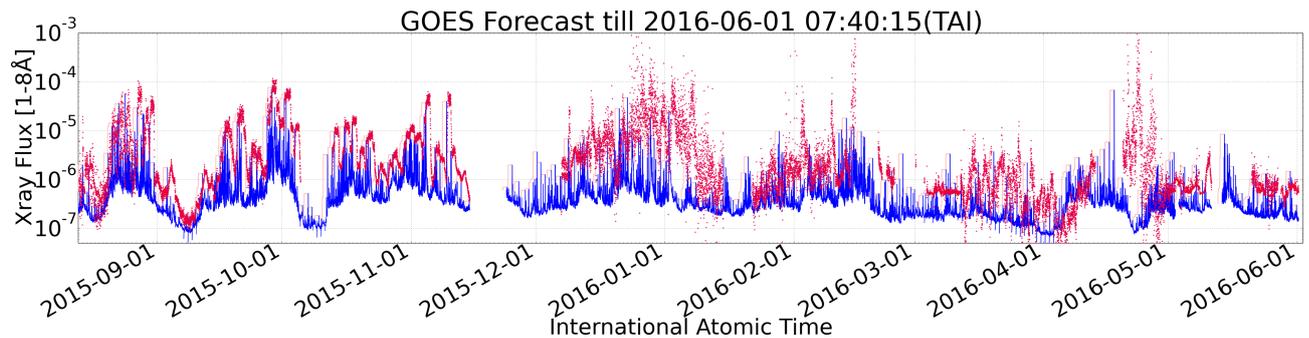

**Figure 4.** Prediction made so far

**Table 7.** The list of system failure events and our action

| begin | end | event | countermeasure |
|---|---|---|---|
| 2015-10-05T11:24 | 2015-10-13T01:36 | Algorithm update | (This was intended pause) |
| 2015-11-15T03:24 | 2015-12-07T03:48 | Data acquisition failure | Implement timeout in JSOC data acquisition script |
| 2016-01-15T02:24 | 2016-01-22T02:00 | System library update | Rewrite UFCORIN source code to the new library interface |
| 2016-02-15T21:36 | 2016-02-21T06:48 | Disk space depletion | Remove unnecessary files |
| 2016-05-11T09:36 | 2016-05-21T01:53 | Hangup of the forecaster | Implement timeout in the forecaster |

One of the possible reason of this discrepancy is the order of the data. In simulated forecast, the predictions are made in the increasing order of time, one per each. The input data to the machine learner is ordered in time. In the real-time forecast, the machine learner repeatedly selects a random 17-day segment from the input time series and learns from it. The order of input data might be an important factor for good TSS. This must be investigated.

Spurious predictions, such as those with $F_x > 10^{-3}$ W/m$^2$, are seen too often in our prediction. At a glance, such spurious predictions have negative effect on the value of the forecast. Re-scaling the prediction by rule-based postprocess may contribute to the increase of the TSS.

The use of wavelet-integral as preprocess greatly limits the types of features that the predictors can recognize. Replacing wavelet filters by convolutional neural networks (CNN) may result in the increase of the TSS.

In any case, any "improvements" to the predictor must be tested and valued in real-time forecast. Our UFCORIN engine helps such planning-testing cycle. Finally, more efforts are desired in decreasing the downtime of such real-time forecast framework.

## Appendix A: Technical Details of The Realtime Forecast Server

The system is operated on Amazon Web Services (AWS). The list of virtual machines is in Table 6.

## Appendix B: Programs and Data Availability

In compliance with AGU's Data Policy, we provide access to the computer programs and the data we have used in this research. The source code for UFCORIN is published at https://github.com/nushio3/UFCORIN. The data is hosted at Amazon S3 (Simple Storage Service); please contact authors for access to the data. The data was originally obtained from SDO/HMI and GOES websites, via URLs http://satdat.ngdc.noaa.gov/sem/goes/data/ and http://jsoc.stanford.edu/.

**Acknowledgments.**
This work was supported by a Grant-in-Aid from the Ministry of Education, Culture, Sports, Science and Technology of Japan (No. 25287039). Part of this research used the computational resources supported by the RIKEN Advanced Institute for Computational Science(AICS).

This is a joint research project of Kyoto University and Broad-Band Tower, Inc. We thank Hiroshi Fujiwara, the CEO of Broad-Band Tower, Inc. for his continuous support and encouragement that made this research possible in the first place. We thank NASA SDO/HMI team and the GOES team for the data used in this study.